\documentclass[12pt]{article}
\usepackage{amsmath,amsthm,amssymb}
\usepackage{graphicx,psfrag,epsf}
\usepackage{enumerate}
\usepackage{natbib}
\usepackage{url} 

\usepackage{lineno,sectsty,ctable,bm,enumerate,booktabs}
\usepackage[title]{appendix}

\newtheorem{theorem}{Theorem}
\newcommand{\blind}{0}

\newcommand{\bEta}{\bm{\eta}}

\begin{document}

\def\spacingset#1{\renewcommand{\baselinestretch}%
{#1}\small\normalsize} \spacingset{1}


\if0\blind
{
  \title{\bf Assessing Biosimilarity using Functional Metrics}
  \author{Sujit K. Ghosh and Lin Dong\hspace{.2cm}\\
    Department of Statistics, North Carolina State University}
\date{}
  \maketitle
} \fi

\if1\blind
{
  \bigskip
  \bigskip
  \bigskip
  \begin{center}
    {\LARGE\bf Assessing Biosimilarity using Functional Metrics}
\end{center}
  \medskip
} \fi

\bigskip
\begin{abstract}
In recent years there have been a lot of interest to test for similarity between biological drug products, commonly known as biologics. Biologics are large and complex molecule drugs that are produced by living cells and hence these are sensitive to the environmental changes. In addition, biologics usually induce antibodies which raises the safety and efficacy issues. The manufacturing process is also much more complicated and often costlier than the small-molecule generic drugs. Because of these complexities and inherent variability of the biologics, the testing paradigm of the traditional generic drugs cannot be directly used to test for biosimilarity. Taking into account some of these concerns we propose a functional distance based methodology that takes into consideration the entire time course of the study and is based on a class of flexible semi-parametric models. The empirical results show that the proposed approach is more sensitive than the classical equivalence tests approach which are usually based on arbitrarily chosen time point. Bootstrap based methodologies are also presented for statistical inference.
\end{abstract}

\noindent%
{\it Keywords:}  Binary Responses; Bernstein Polynomials; Rheumatic Arthritis; Semiparametric Models
\vfill
\newpage

\spacingset{1.45} 
\section{Introduction}
\label{intro}
Biosimilars is referred to as the similarity between biological drug products. Biologics are large and complex molecule drugs that are produced by living cells. They have heterogeneity structures and are sensitive to the environmental changes. In addition, biologics usually induce antibodies which raises the safety and efficacy issues. The manufacturing process is also much more complicated and costly than the small-molecule generic drugs. Due to such nature of biologics, the assessment of biosimilar products is fundamentally different from the generic small-molecule drug products. With the small-molecule drug, the chemical formula is normally known and can be recreated exactly the same as the innovator drug, hence randomized clinical trials are routinely not required for the approval. Because of the complexity and the variability of the biologics, the paradigm of the traditional generic drugs cannot be directly extended to biosimilars. Further details on biosimilars can be seen in guidance \cite{Fda_Bios:2012}, and statistical literatures such as \cite{Lia+Hey:2011}, \cite{End+Cha+Cho+Tot:2013} and \cite{Cho:2013}.

To establish the biosimilarity between the test drug and the innovator, a biosimilar manufacturer needs to show the equivalence in both of the pharmacokinetic (PK) and pharmacodynamic (PD) parameters, and the therapeutic effects. The equivalence of PK parameters, such as the area under the curve (AUC) and the maximum observed concentration (Cmax), can be determined similarly to the generic drugs using the classic average bioequivalence rule, in which bioequivalence can be claimed when the 90\% confidence interval for the estimated ratio of geometric mean (GMR) of the parameters lies between the limits 0.80 and 1.25 (see \cite{FDA:2001}). This classic bioequivalence rule can also be challenging for the biologics to pass, as biologics are far more variable than usual small molecule drugs. Some reference scale approaches for biosimilarity which take the reference variability into consideration have been proposed by \cite{Hai+etal:2008b}, \cite{Hai+etal:2008},  and \cite{Lia+Hey:2011}. However, it is more challenging when it comes to the equivalence based on clinical endpoints, and no statistical approach has been well-defined.

Classically, the equivalence of therapeutic effects can be declared if the difference between the means of the primary endpoints for two drugs is less than a pre-specified non-inferiority limit, with 95\% confidence. However, for biological drugs, this approach can result in large sample size, which is against the goal to reduce the expense of the biosimilar product. Moreover, the biosimilarity based on the primary endpoints is only established at one single time point, typically when the maximal effect is achieved by the innovator. However, it could be possible that differences can be observed at early, non-saturated time points. Such differences before the primary time point cannot be captured by this approach.

To overcome this disadvantage, we propose a functional distance based methodology that takes into consideration of the entire time course. Although our proposed methodology is applicable to many types of diseases and similar scenarios, we illustrate our methodology for the disease rheumatoid arthritis (RA) as a case study. In recent years, treatment of RA has gained increasing interest in the pharmaceutical industry. Many biologics have been developed and used in clinical trials including etanercept, adalimumab (HUMIRA), infliximab, golimumab, tocilizumab, abatacept, and certolizumab, some of these have generated multi-billion revenue per year for the pharmaceutical companies. As a result of large profit, there is large interest in developing innovator biological drugs, and biosimilar drugs. 

In Section 2, we introduce the proposed functional metrics and present two classes of modeling techniques. Section 3 is devoted to estimation methods for fitting such models and thus estimators for functional metrics. In Section 4,  we explore several simulation experiments to investigate the performance of these estimators. In Section 5, we apply the proposed functional metrics to various Rheumatoid Arthritis Trials data sets for illustration. Finally in Section 6, we make some concluding remarks and discussion on possible extensions.

\section{Functional Metrics to Assess Biosimilarity}
\label{s:methodology}
In a treatment group with $n_j$ subjects, let $Y_j(t)$ denote the number of subjects that achieves some clinical event at time $t$ for group $j = 1,\dots,J$. Assuming that the trials were performed independently for each group, we postulate the familiar binary response model: 
\begin{equation}\label{eq:y}
Y_j(t) \sim \mathrm{Bin}\{n_j,\theta_j(t)\}, \text{for }t\ge0,
\end{equation}
where $\theta_j(t)$ is the response rate function for group $j = 1,\dots,J$.

When there are only $J = 2$ groups, we write the (additive) difference of response rates as a function of time $\Delta(t) = \theta_2(t) - \theta_1(t)$. Notice that the difference in rates is allowed to vary with time $t$ in contrast to traditional approach which usually chooses a specific time point $t_0$ and uses fixed difference $\Delta_0=\Delta(t_0)$ to decide the difference between two response rates at a fixed time point $t=t_0$. Many other discrepancies between these rates can also be used (e.g., relative difference, odds ratio or log odds ratio), if so desired and the proposed methodology can easily be extended to such alternative (time varying) measures of discrepancies. 

Taking a more dynamic approach, we propose a functional metric $L_p(a,b)$ that takes into account the entire time interval to compare the groups at a chosen interval of time. We define the functional metric as follows:
\begin{equation}
   L_p(a,b) = \left(\int_a^b \vert \Delta(t) \vert^p dt\right)^{\frac{1}{p}},\quad p\ge 1\text{ and }a,b \subseteq [0,\infty).
\end{equation}
In practice, it is common to choose $p = 1,2$ and $\infty$, representing $L_1, L_2$ and $L_\infty$ norms. Notice that $L_\infty(a,b) = \underset{a\let\le b}{\mathrm{max}}\vert \Delta(t)\vert$. Although the user needs to choose a sub-interval $(a, b)$ based on the therapeutic considerations for a specific drug and trail, one can mitigate the effect to some effect by considering the scaled metric $ L_p(a,b)/(b-a)$, which would remain relatively stable with respect to the interval length $b-a$.
For common dose-response relationship, it is often assumed that the true response rate function $\theta_0(t)$ satisfy the following conditions:
\begin{enumerate}[(i)]
    \item $\theta_0(\cdot)$ is continuous in $[0,\infty)$; \label{cond:continuous}
    \item $\theta_0(0) = 0,$ and $\theta_0(\infty) \le 1$;\label{cond:rate}
    \item (optional) $\theta(t_1)\le \theta(t_2), \text{for } 0\le t_1\le t_2 \le \infty$.\label{cond:mono}
\end{enumerate}
Condition (\ref{cond:mono}) is the monotonic shape constraint on the entire time interval, which could be dropped or relaxed depending on type of clinical trials and associated drugs under such studies. Our goal is to obtain estimate of the functional metric by suitably modeling the true response rates for each group as flexibly as possible while satisfying the above mild regularity conditions.

Given a chosen value for the triplet $(p,a,b)$, often the goal is to develop statistical inference procedure to test the following non-inferiority hypotheses
\begin{align*}
H_0: L_p(a,b) > d(p;a,b)\;\;\;\mbox{vs.}\;\;H_A: L_p(a,b) \le d(p;a,b),
\end{align*}
where the non-inferiority margin $d(p; a,b)$ is determined by medical practitioners to achieve the desired power of the test. 

\noindent {\em Remark: We would like to point out that our focus is on deriving statistical inference procedures for the proposed functional metric $L_p(a,b)$ given a specific non-inferiority margin and not necessarily in determining the margin.}

Nonetheless, the choice of the margin is a critical matter and should be carried out very carefully by consulting with the regularity agencies and drug makers. Our inferential procedures described later, does not depend on a any specific choices of this margin and can be broadly applied to any chosen values of the triplet and the corresponding margin.

In the following subsections, we introduce two classes of models to estimate $L_p(a,b)$: parametric models and nonparametric models.

\subsection{Parametric Models}
There are a multitude of choices to model the dose response function $\theta(t)$, satisfying the above mentioned regularity conditions and based on the type of applications, here, however we present two popular class of parametric models that are routinely used for RA trials. 

A popular and very useful class is given by the so-called {\em exponential decay} model given by the following equation:
\begin{equation}\label{e:exp}
\theta_E(t) = \alpha\left\{1-\mathrm{exp}(-\beta t)\right\},\quad t>0,
\end{equation}
where $\alpha$ and $\beta$ are unknown positive quantities to be estimated (see  \cite{reeve2013rheumatoid} for further details). For $\theta_E(\cdot)$ to satisfy the above three properties, we have to restrict $\alpha \in (0,1]$ and $\beta > 0$. We refer model form (\ref{e:exp}) as the exponential decay model. Notice that $\theta_E(t)$ is a strictly increasing function of $t$ and has an asymptote at $\theta_E(\infty)=\alpha$ and hence the restriction $\alpha\leq 1$.

\begin{figure}[h!]
    \centering
    \includegraphics[width = 0.5\textwidth]{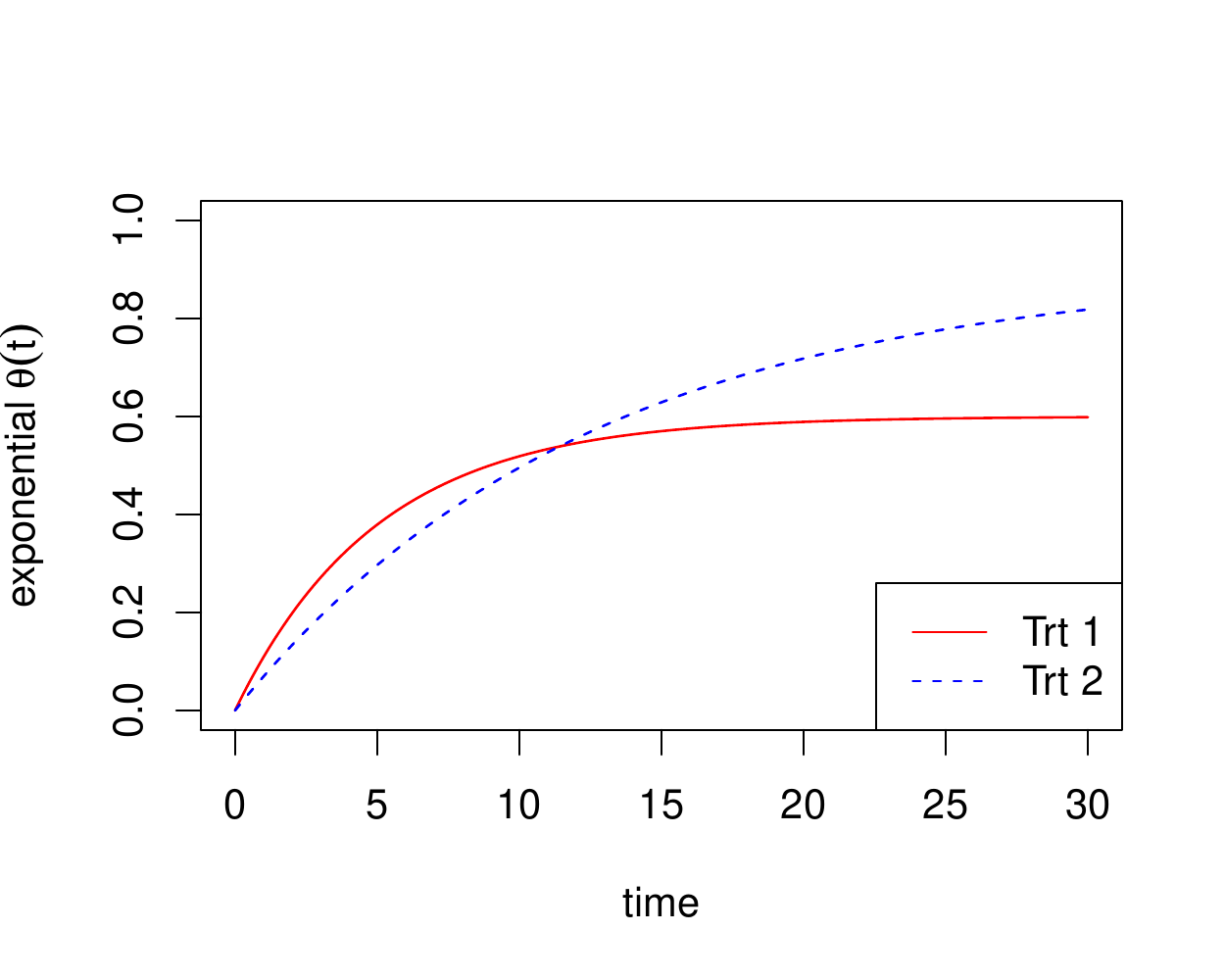}
    \caption{A demonstration of exponential decay curves with $\alpha_1 = 0.6, \beta_1 = 0.2$ and $\alpha_2= 0.9, \beta_2 = 0.08$.}
    \label{fig:expdemo}
\end{figure}

\begin{figure}[h!]
    \centering
    \includegraphics[width = 0.5\textwidth]{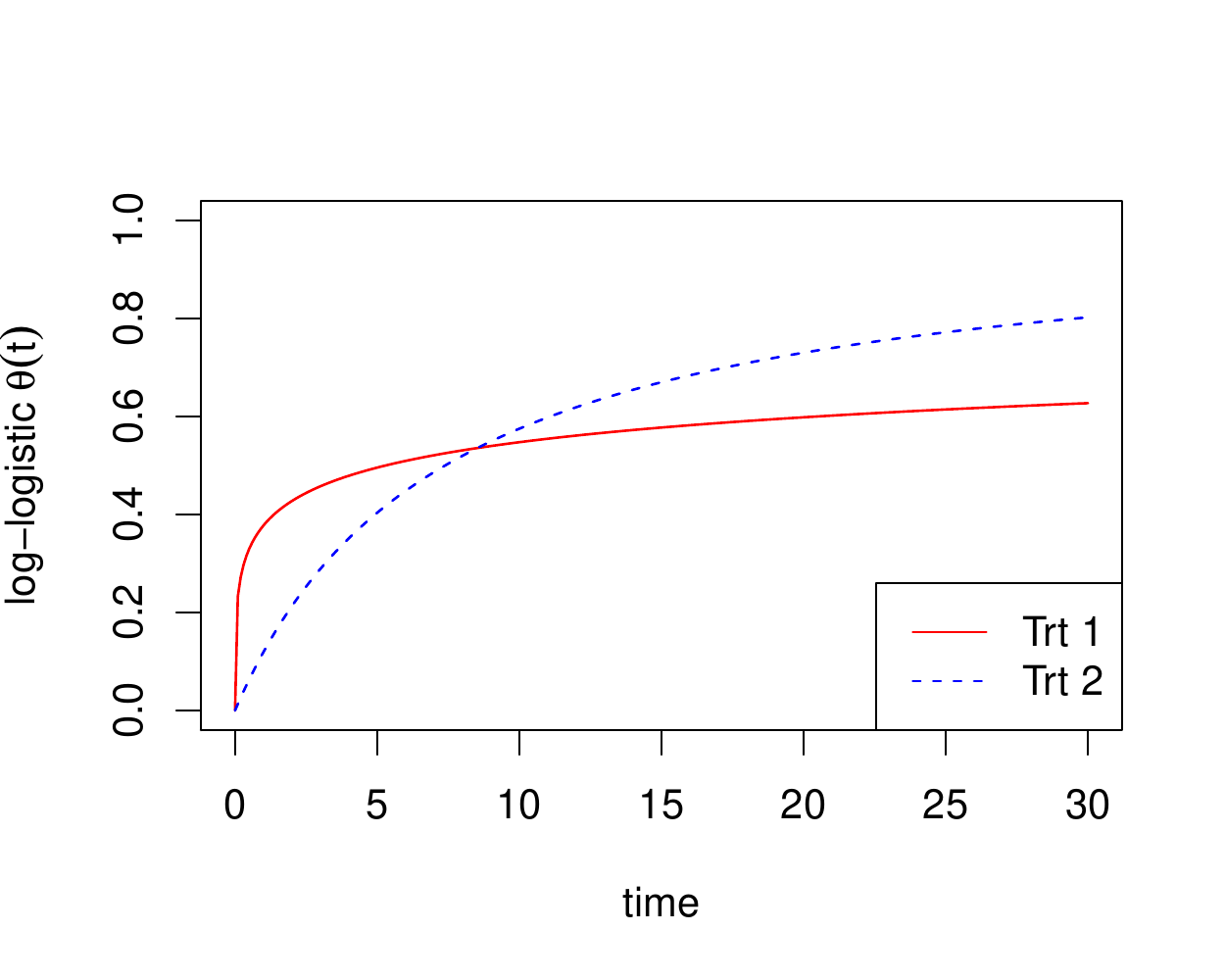}
    \caption{A demonstration of log-logistic curves with $\alpha_1 = -0.5, \beta_1 = 0.3$ and $\alpha_2= -2, \beta_2 = 1$.}
    \label{fig:loglogdemo}
\end{figure}

Another parametric form that is also used very frequently is the so-called {\em log-logistic} model given by,
\begin{equation}\label{e:loglog}
    \theta_L(t) = \left( 1 + e^{-\alpha - \beta \mathrm{log}t}\right)^{-1}.
\end{equation}
$\theta_L(\cdot)$ also has two unknown parameters and satisfies the three properties for any $\alpha \in \mathbb{R}$ and $\beta \ge 0$. Notice that $\theta_L(t)$ is also a strictly increasing function with an asymptotic $\theta_L(\infty)=1$. This model has a similar shape compared with the exponential decay model as shown in Figure~\ref{fig:expdemo} and Figure~\ref{fig:loglogdemo}.

In subsequent parametric likelihood based analysis, we will assume that the response rate functions are modeled using either exponential decay model or the log-logistic model and then we can compute the difference $\Delta(t)$ using numerical integration methods (e.g., Gaussian quadrature methods using the R function {\tt integrate}) given a value of the two parameters $(\alpha, \beta)$ in either cases, be it for the true value or estimated values of these parameters, when performing simulation studies or analyzing case studies, respectively.

\subsection{Nonparametric Models}
A well-known shortcoming of using parametric models is that specifying an incorrect parametric form leads to bias and inefficient estimation. Therefore, a more flexible functional form is desirable as the true underlying treatment effect model is most likely unknown. Bernstein polynomial basis can be used in function estimation with shape restriction. A Bernstein polynomial of degree M is given by
\begin{equation} \label{e:Bern}
B_M(x,\bm{\eta}) = \sum_{k=0}^{M}\eta_k b_M(x,k),
\end{equation}
where $b_M(x,k) = \binom{M}{k} x^k(1-x)^{1-k}$, which is the binomial probability mass function or equivalently can be viewed as the kernel of a Beta (distribution) density function. 

Let $T_\mathrm{min}$ and $T_\mathrm{max}$ be the time range with $0\le T_\mathrm{min} < T_\mathrm{max} <\infty$. We consider the following functional form (see \cite{shin2017comparative} for details).
\begin{equation}\label{e:bern}
\theta_M(t;\bm{\eta},T_\mathrm{max},T_\mathrm{min}) = 
\begin{cases}
0, \quad t \le T_\mathrm{min}\\
\sum_{k = 1}^{M} \eta_k b_M\left(\frac{t-T_\mathrm{min}}{T_\mathrm{max} - T_\mathrm{min}}, k\right), \quad T_\mathrm{min} < t < T_\mathrm{max}\\
\eta_M + (1 - \eta_M)\frac{t - T_\mathrm{max}}{t - T_\mathrm{max} + 1}, \quad t\ge T_\mathrm{max}
\end{cases}
\end{equation}
with the constraint 
\begin{equation*}
0\le\eta_1\le\dots\le\eta_M\le1.
\end{equation*}

A computational convenient reparametrization of (\ref{e:bern}) is to define $\eta_0 = 0$ and $\gamma_k = \eta_k - \eta_{k-1}$, so that $\eta_k = \sum_{i=1}^k\gamma_i$ and (\ref{e:Bern}) can be re-write as
\begin{equation*} 
B_M(x,\bm{\eta}) = \sum_{k=1}^{M}\eta_k b_M(x,k) = \sum_{k=1}^M \left(\sum_{l=1}^k\gamma_l\right) b_M(x,k) = \sum_{l=1}^M \gamma_l F_M(x;l),
\end{equation*}
where $F_M(x;l) = \sum_{k=l}^{M}b_M(x,k)$.
The benefit of this reparametrization is that we now express $\theta_M(t)$ as a linear model:
\begin{equation*}
    \theta_M(t;\bm{\gamma}) = \bm{F}_M\left(\frac{t-T_\mathrm{min}}{T_\mathrm{max} - T_\mathrm{min}}\right)^\tau\bm{\gamma},
\end{equation*}
where $ \bm{F}_M(x) = (F_M(x,1),\dots,F_M(x,M))^\tau$ denotes vector of basis functions. A convenient fact about $F_M(x;l)$ is 
\begin{equation*}
    F_M(x;l) = \int_0^x \frac{u^{l-1}(1-u)^{M-l}}{B(l,M-l+1)}du,
\end{equation*}
which is the cumulative density distribution of a $\mathrm{Beta}(l,M-l+1)$ random variable. The parameter $\bEta$ can be estimated by maximum likelihood methods, which we will discuss in section \ref{s:approxLik}. 

A well known fact about the Bernstein basis is that it provides uniform approximation of a continuous function on a compact interval. More specifically, by choosing $\eta_k=\theta_0(k(T_{max}-T_{min})/M)$ in equation (\ref{e:bern}) one can show that $\sup_{t\in [T_{min}, T_{max}]}|\theta_M(t;\bm{\eta},T_\mathrm{max},T_\mathrm{min})-\theta_0(t)|\rightarrow 0$ as $M\rightarrow\infty$ (see \cite{lorentz2012bernstein}). Rate of convergence can also be obtained by assuming further regularity conditions on the true understing dose response function $\theta_0(t)$. These well known results together with the sieve based estimation techniques presented in \cite{geman1982nonparametric} and \cite{shen1994convergence} can be used to derive the consistency of our proposed methods described in the next section. However, we omit the technical details in this paper.

\section{Estimation Methods}\label{s:Method}

\subsection{Maximum Likelihood for Parametric Models}
Once we specified the functional form of $\theta(t)$, we can get maximum likelihood estimates (MLEs) of the model parameters $(\alpha
_j, \beta_j)$ for each treatment arm by maximizing the log-likelihood
\begin{equation}\label{likelihood}
    l(\alpha_j,\beta_j) = \sum_{i=1}^{N_j}Y_j(t_i)\mathrm{log}(\theta(t_i;\alpha_j,\beta_j)) + (n_j - Y_j(t_i))\mathrm{log}(1 - \theta(t_i;\alpha_j,\beta_j)),\quad j = 1,2,
\end{equation}
where $N_j$ is the number of time points that we have observations for each treatment arm $j$.
The estimated treatment effect curve is obtained by plugging in the maximum likelihood estimates of parameters for each treatment arm,
\begin{equation*}
\widehat{\Delta}(t;\hat{\alpha}_1,\hat{\beta}_1,\hat{\alpha}_2,\hat{\beta}_2) = \theta(t;\hat{\alpha}_2,\hat{\beta}_2) - \theta(t;\hat{\alpha}_1,\hat{\beta}_1).
\end{equation*}

The proposed function metric $L_p(a,b)$ is then estimated by plugging in the estimated treatment effect curve,
\begin{equation*}
   \hat{L}_p(a,b;\hat{\alpha}_1,\hat{\beta}_1,\hat{\alpha}_2,\hat{\beta}_2) = \left(\int_a^b \vert\widehat{\Delta}(t;\hat{\alpha}_1,\hat{\beta}_1,\hat{\alpha}_2,\hat{\beta}_2) \vert^p dt\right)^{\frac{1}{p}},\quad p\ge 1\text{ and }a,b \subseteq [0,\infty),
\end{equation*}
which can be easily computed by numerical integration method. The corresponding estimates of the standard errors of the MLEs and hence the associated confidence intervals can be obtained using the standard so-called delta-method using standard large sample theory (e.g., see \cite{casella2002statistical}). But such methods can lead to complicated expression and may require numerical derivative and integration calculations. Alternatively, we use the standard parametric bootstrap methods to derive the sampling distributions of the estimates of $\Delta(t)$ and hence obtain the standard errors and confidence intervals.

\subsection{Approximate Likelihood Methods for  Nonparametric Models}\label{s:approxLik}

Similar to parametric models, the estimation can be done by maximizing the log-likelihood (\ref{likelihood}) where we substitute $\theta(\alpha,\beta)$ by $\theta_M(\bEta)$ for a fixed $M$ and then vary $M$ as a function of the sample size $n$ to derive the nonparametric estimate of $\Delta(t)$. However, such a method can be computationally demanding due to nonlinear optimization subject to linear inequality constraints on $\bEta$. 

An alternative method, which is more computationally efficient, is to use the normal approximate likelihood methods, when sample size $n$ is moderately large. Using the standard normal approximation of the Binomial likelihood, it follows that the distribution of estimated response probability $\theta(t)$ can be approximated by the following normal distribution
\begin{equation}\label{appr_like}
    \sqrt{n_j}\left\{\hat{\theta_j}(t) - \theta_j(t)\right\} \sim N\left[0,\theta_j(t)\{1 - \theta_j(t)\}\right], j = 1,2,
\end{equation}
where $\hat{\theta_j}(t) = Y_j(t)/n_j$ is the empirical nonparametric estimate of $\theta_j(t)$. In the case of $Y_j(t) = 0$, one can use Anscombe corrected empirical proportion given by $\hat{\theta_j}(t) = \frac{Y_j+3/8}{n_j+3/4}$ which has been shown to have second order accuracy in estimating the true response rate function. 
To account the non-constant variance of the normal distribution, we will adopt a weighted least square method by replacing the true variance by its empirical estimate. The estimation problem now can be cast as a standard  linear regression model, where $\hat{\theta_j}(t)$ serves as the response variable and $\bm{F}_{Mj}(t)$ serves as the predictors.
The estimation of parameters $\bm{\gamma}_j$ is now turns into a constrained weighted least square problem
\begin{align*}
& \underset{\bm{\gamma}_j}{\mathrm{min}}\sum_{i=1}^{N_j}w_{ij} \left\{\hat{\theta}_j(t_i) - \bm{F}_{Mj}(t_i)^\prime\bm{\gamma}_j\right\}^2\\
& \text{subject to} \quad \bm{R} \bm{\gamma}_j \ge \bm{b}, \; j = 1,2.
\end{align*}
The weight $w_{ij}$ can be estimated by $\hat{w}_{ij} = n_j/\left[\hat{\theta_j}(t_i)\{1-\hat{\theta_j}(t_i)\}\right]$ for $j = 1,2$.
To satisfy the properties in section \ref{s:methodology}, a specific configuration is to let
$$
\bm{R} = 
\quad \begin{pmatrix}
\bm{I}_M\\
-\bm{1}_M^\prime
\end{pmatrix}\bm{\gamma}_j \;\text{ and }
\bm{b} = 
\begin{pmatrix}
\bm{0}_M\\
-1
\end{pmatrix}.
$$
This optimization problem can be easily solved by any standard quadtratic programming techniques (e.g., using {\tt quadprog} package in the R software) and hence obtain computationally efficient estimates of $\widehat{\bm{\gamma}}_j$ and hence that of $\widehat{\bm{\eta}}_j$. Recall that, all of these estimation can be done relatively easily for a fixed chosen value of the tuning parameter $M$ which required to increase with the sample size $n$.

This key issue of the selecting the degree of  Bernstein polynomial, $M$ is tricky and we propose a practical approach based on suitable metric to choose $M$ which depends on data. We formally propose a method to make this selection in section \ref{s:selectM}. For now, we assume the $M$ values are pre-determined using the metric.

Once suitable $M$ is determined separately for both treatment groups, we get functional estimates $\widehat{\theta}_1(\cdot;\widehat{\bm{\gamma}}_1) = \theta_{M_1}(\cdot;\widehat{\bm{\gamma}}_1)$ and $\widehat{\theta}_2(\cdot;\widehat{\bm{\gamma}}_2) = \theta_{M_2}(\cdot;\widehat{\bm{\gamma}}_2)$. Similar as parametric models, the plugged in estimator of functional metric $L_p(a,b)$ is
\begin{equation*}
   \hat{L}_p(a,b;\widehat{\bm{\gamma}}_1,\widehat{\bm{\gamma}}_2) = \left(\int_a^b \vert\widehat{\theta}_{2}(t;\widehat{\bm{\gamma}}_1) - \widehat{\theta}_{1}(t;\widehat{\bm{\gamma}}_2)  \vert^p dt\right)^{\frac{1}{p}},\quad p\ge 1\text{ and }a,b \subseteq [0,\infty).
\end{equation*}

\subsection{Selection of $M$ via Kolmogorov-Smirnov metric}\label{s:selectM}
We provide a principled way to select the Bernstein polynomial degree $M$ from the observed data. The idea here is to use the Kolmogorov-Smirnov (KS) metric to make data-driven decision for polynomial degree selection. Notice that we have used the normal approximation of empirical estimates of the response rate function to derive the estimate of $\bEta$ and so if an $M$ is selected that makes the normal approximation superior across different choices of $M$, the estimate of $\bEta$ will also likely to be better in approximating the true response rate function.

Using the above intuitive principle, given the approximate likelihood given in (\ref{appr_like}), we can construct a vector of standardized residuals which are likely to be $N(0,1)$ distributed variables
$$
Z_i(m) = \frac{\sqrt{n}\left\{\hat{\theta}(t_i) - \hat{\theta}_m(t_i)\right\}}{\theta(t_i)\{1 - \theta(t_i)\}}, i = 1,\dots,N,
$$
where $n$ is sample size, $N$ is the number of time points and $\hat{\theta}_m(t_i)$ is the fitted value obtained from a Bernstein polynomial of degree $m$.
We want to find the smallest $m$ such that the resulting $Z_i(m)$ are as closely approximately normally distributed as possible. Recall that here we are not trying to test for normality, but rather using the KS metric to select `optimal' degree $m$ that minimizes the KS metric over a given set of values of $m$. As the KS metric can take on any positive value, we use the equivalent $p$-value of the KS test which is automatically scaled to be between $0$ and $1$. Clearly, the closer the $p$-value is to $1$ the better the standardized residuals match with a $N(0, 1)$ distribution. Therefore, we we compute the $p$-value of the KS test for $m = 2,\dots,\lceil\frac{N}{\log(N)}\rceil$ (\cite{babu2002application}) and select $\hat{M}_\mathrm{opt} = m$ whenever the p-value of the test is greater or equal to our pre-specified critical value $\alpha\in (0, 1)$. A default value for $\alpha=0.2$ is reasonable in practice but we explore its sensitivity using simulation studies in selecting the optimal $M$. 

\section{Simulation Study}
\label{simul}

In this section, the performance of the proposed function metrics are investigated under several functional forms using simulated data sets. We explore and compare two categories of models, namely parametric models and nonparametric models. All results in this section are based on 1000 Monte Carlo repetitions. 

\begin{figure}[h]
    \centering
    \includegraphics[width = 0.6\textwidth]{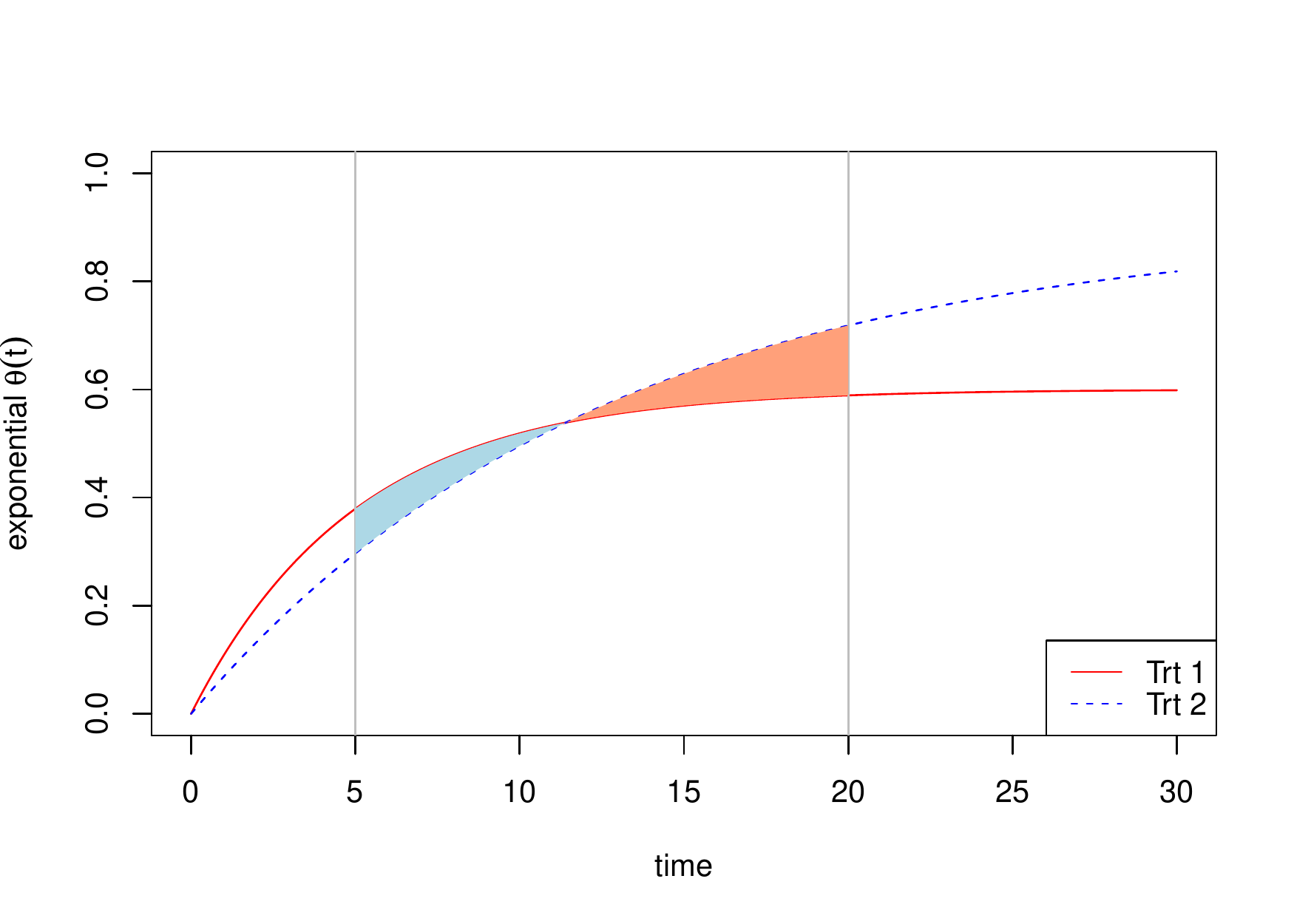}
    \caption{Response rate curves of the two groups used for simulation. The $L_1(5,20)$ quantity is the orange shaded area plus the blue shaded area.}
    \label{fig:true}
\end{figure}

We simulate two treatment group by the exponential decay model (\ref{e:exp}) with $\alpha_1 = 0.6, \beta_1 = 0.2$ and $\alpha_2 = 0.9, \beta_2 = 0.08$ respectively. The time horizon is set to be $[0,30]$. We simulate an observation of the response rate in every $1$ or $2$ time units. Therefore, the number of observations we have for each treatment group is $N = 16$ or $N = 31$ respectively. To evaluate functional metric $L_p(a,b)$, time boundary $a=5, b = 20$ and $L_1$ norm ($p = 1$) are used through out this section. This generative model and the corresponding $L$ value is demonstrated in Figure~\ref{fig:true} and these choices are motivated by the real case studies that we illustrate in the next section.

We compare the performance of parametric model and nonparametric model with varying sample sizes and varying number of time points settings for estimating the true functional metric quantity $L_0 = L_1(5,20)$. For parametric model, the correctly specified model is exponential decay model (\ref{e:exp}). As we are interested in exploring the effect of misspecification of the underlying true model,  a log-logistic model (\ref{e:loglog}) is fitted for comparison (when the true underlyting model is the exponential decay model), which we refer to as the {\em misspecified parametric (MP)} model. Nonparametric model is implemented by the approximate likelihood method introduced in section \ref{s:approxLik}. The polynomial degree is selected according to the method introduced in section \ref{s:selectM}.

\begin{figure}[h!]
\centering
  \includegraphics[width=\linewidth]{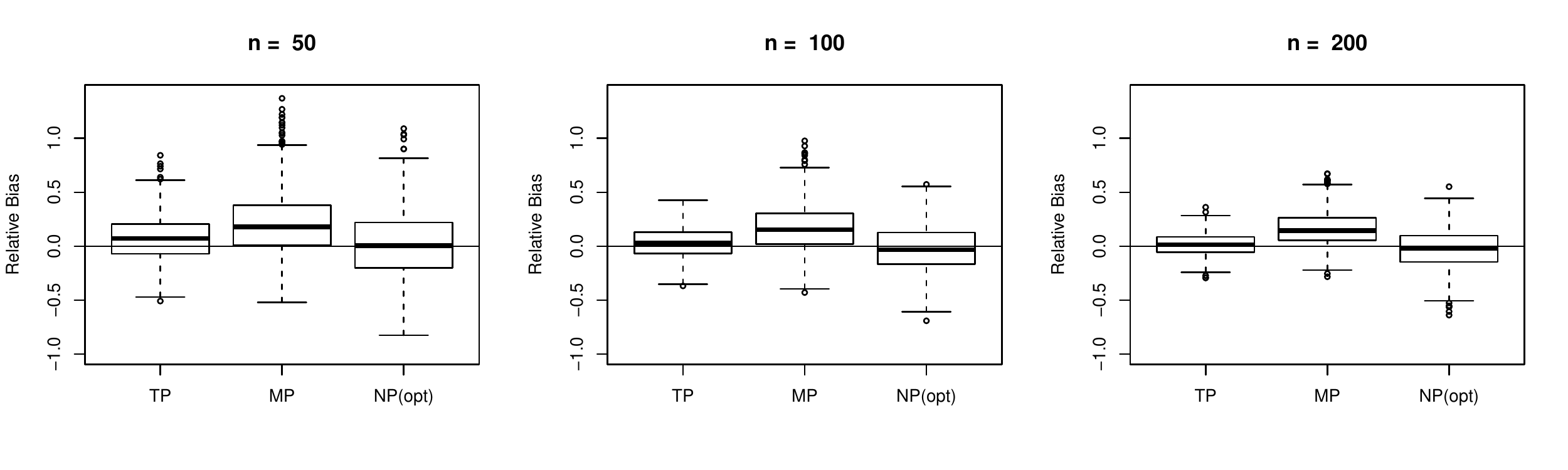}
    \caption{Boxplot of relative bias. N = 16} 
        \label{f:bias_n_N16}
      \includegraphics[width=\linewidth]{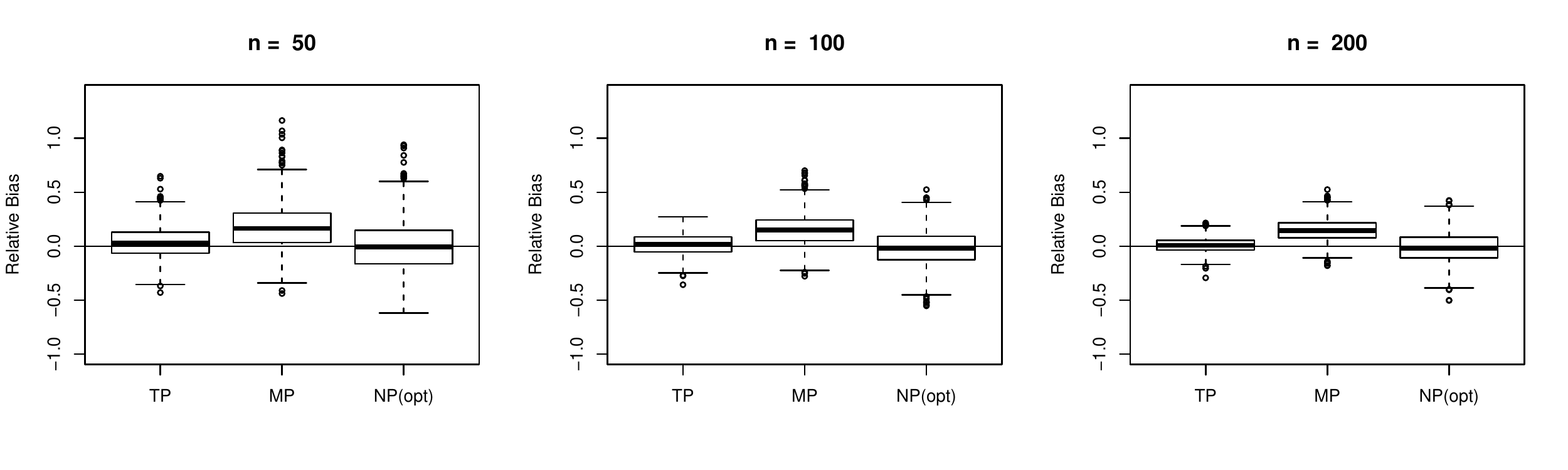}
    \caption{Boxplot of relative bias. N = 31 } 
    \label{f:bias_n_N30}
\end{figure}

To compare performance of the proposed estimators, we define the {\em relative bias (RB)} of estimator $\hat{L}$ as $\mathrm{RB}(\hat{L}) = (\hat{L} - L_0)/L_0$, where $L_0$ is the true value of the $L_1$ metric. The results comparing the values of RB of $\hat{L}$ under different functional forms are reported in Table~\ref{t:sim_M_n}. Boxplots of RBs (across 1000 MC runs) are shown in Figure~\ref{f:bias_n_N16} and Figure~\ref{f:bias_n_N30}. A summary of the selected values of M  is reported in Table~\ref{t:meanMopt}.

Note that the log-logistic model (\ref{e:loglog}) has very similar shape compared to the exponential model (\ref{e:exp}); however, the parametric estimator using log-logistic model still demonstrates a high notable bias (in terms of RB) as can be seen in Table~\ref{t:sim_M_n}.
Nonparametric methods are shown to give relatively unbiased estimates of the functional metric, and have competitive performance comparing it to the correctly specified parametric model. As expected when the sample size $n$ and number of sampling points $N$ increases, the RB of the $L$ metric under MP tends to increase while those under TP and NP shrinks. This is a well-known phenomenon of using MLE under misspecified models and is clearly depicted in our experimental simulation studies. 

\begin{table}[h!]
\centering
\begin{tabular}{|c|ccc|ccc|}
  \hline
 & \multicolumn{3}{c|}{N = 16} & \multicolumn{3}{c|}{N = 31}\\
  \hline
 n & TP & MP & NP(opt) & TP & MP & NP(opt) \\ 
  \hline
 50  & 0.073 & 0.209 & 0.014 & 0.035 & 0.178 & 0.001 \\ 
     & (0.0065) & (0.0094) & (0.0097) & (0.0046) & (0.0069) & (0.0075) \\ 
 100 & 0.030 & 0.168 & -0.029 & 0.016 & 0.154 & -0.018 \\ 
     & (0.0045) & (0.0067) & (0.0069) & (0.0033) & (0.0048) & (0.0053) \\ 
 200 & 0.017 & 0.163 & -0.026 & 0.009 & 0.149 & -0.009 \\ 
     & (0.0033) & (0.0049) & (0.0058) & (0.0022) & (0.0033) & (0.0042) \\ 
  \hline
\end{tabular}
\caption{
Relative bias (Monte Carlo standard error) of estimating the functional metric $L$ under various values of $n$ and $N$. TP refers to true (correctly specified) parametric models; MP refers to misspecified parametric model. NP refers to non-parametric approximate normal likelihood method with a Bernstein polynomial of degree M, where M is selected according to the optimal strategy.}\label{t:sim_M_n}
\end{table}

\begin{table}[h]
\centering
\begin{tabular}{|c|cc|cc|cc|cc|}
  \hline
 & \multicolumn{4}{c|}{N = 16} & \multicolumn{4}{c|}{N = 31}\\
  \hline 
 n& \multicolumn{2}{c|}{Treatment 1} & \multicolumn{2}{c|}{Treatment 2}& \multicolumn{2}{c|}{Treatment 1}& \multicolumn{2}{c|}{Treatment 2}\\
 \hline
 50  & 4.867&(0.0399)& 3.590&(0.0405)& 5.585&(0.0660)& 3.903&(0.0710)\\ 
 100 & 5.200&(0.0338)& 3.601&(0.0402)& 5.824&(0.0596)& 3.793&(0.0641)\\ 
 200 & 5.446&(0.0294)& 3.657&(0.0407)& 6.140&(0.0568)& 3.868&(0.0667)\\ 
 \hline
\end{tabular}
\caption{ Selected value of M (Monte Carlo standard error) in the NP(opt) model, with critical value $\alpha = 0.5$. }\label{t:meanMopt}
\end{table}

\section{A Case Study: RA Trials}
\label{appln}
The development of our models has been motivated by the Rheumatoid Arthritis (RA) trails and so an illustration, we apply the proposed method on the Rheumatoid Arthritis (RA) Trials data set (\cite{reeve2013rheumatoid}). The authors collected data from a list of available literature of randomized, double-blind clinical studies in RA, where all the studies used methotrexate (MTX) as the baseline treatment. We choose ACR20 as the endpoint, which is among the most commonly used endpoints for RA. As we do not conduct meta analysis in this paper, we select studies based on the following rule: for all the treatments (molecule $\times$ dose) that have more than 40 weeks of observation, we select the study with the largest number of time points. There are 3 treatment selected from the RA data set  to demonstrate our methods, namely infliximab dose 3 (infilix:3), tocilizumab dose 4(tocili:4) and czp dose 200 (czp:200). As there were several studies that utilized the baseline MTX we report a meta-analytic analysis in the Appendix of the paper that uses random coefficient models and an over estimated response rate curve for the MTX is reported.

We fit both parametric exponential decay model (\ref{e:exp}) and non-parametric Bernstein polynomial model (\ref{e:bern}) on the given set of data. For parametric fit, parameter estimates along with their standard errors are reported in table \ref{t:real_par}. When we apply the non-parametric fit on the data, we found that using the strict monotonic Bernstein polynomial, treatment inflix:3 doesn't obtain an M value that will exceed $\alpha =0.2$ (The largest p-value achieved for this trial is 0.138). To resolve this somewhat poor fit, one strategy is to use data augmentation. If we are willing to assume that at week 0 the response rate should be 0, we can add data point $(0, 0)$ to the original data. Note that most of the treatment do not have week 0 as observed value. Then we treat it as if we observed (0,0) and apply the monotonic fit. Another practical strategy is to partially relax the monotonic constraint (iii) as stated in Section 2. Say, if we allow the Bernstein polynomial to move freely within the first 1/3 knots, the monotonic constraint remains for the last 2/3 of the  knots. Using either of these two strategies, we can have an optimal $M$ selected that would exceed $\alpha = 0.2$ significance level. We use the later strategy as we the strict monotonicity may not be required for all dose response models.

\begin{figure}[ht]
\centering
  \includegraphics[width=0.9\linewidth]{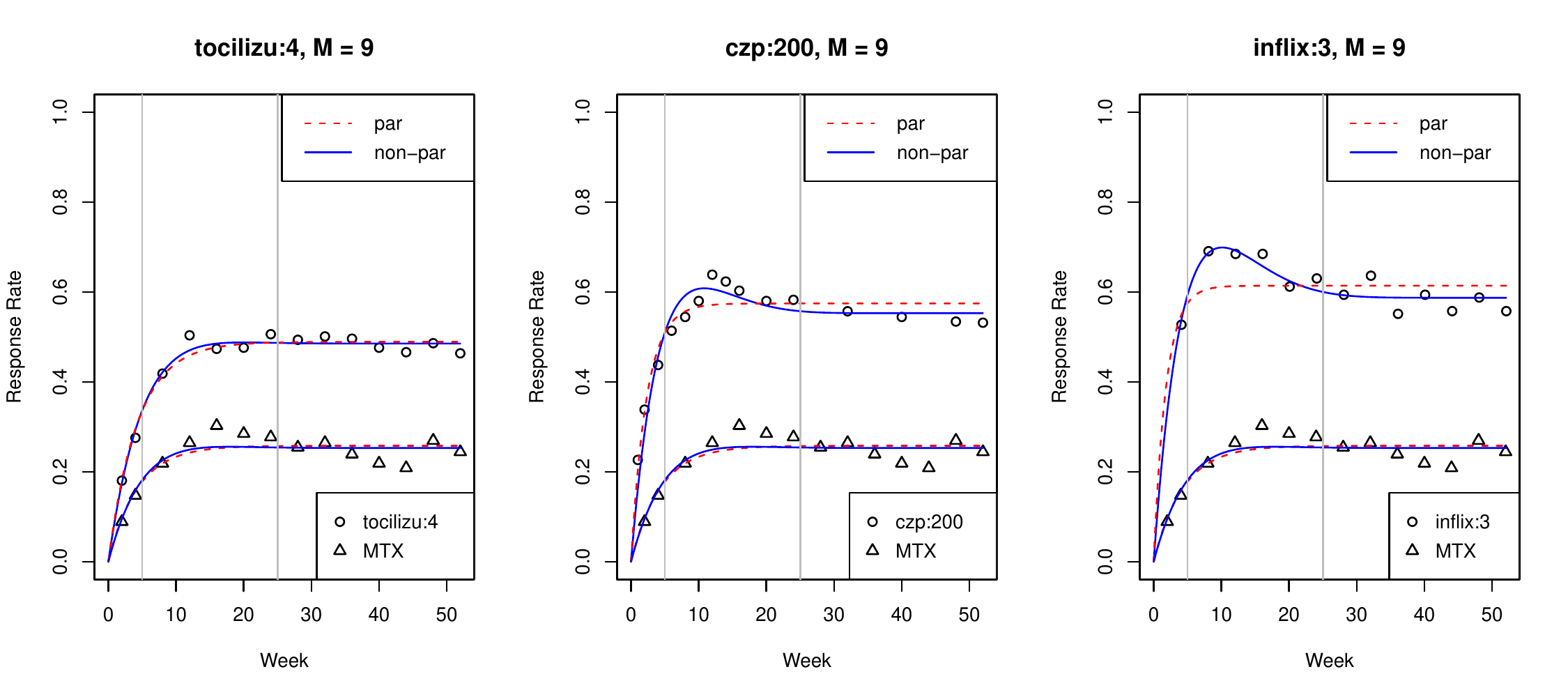}
    \caption{Parametric and nonparametric fit of data by treatment, with M selected based on $\alpha = 0.2$. Grey vertical line corresponding to time interval (5, 25). }\label{f:realdata4}
\end{figure}

\begin{figure}[h!]
\centering
  \includegraphics[width=0.8\linewidth]{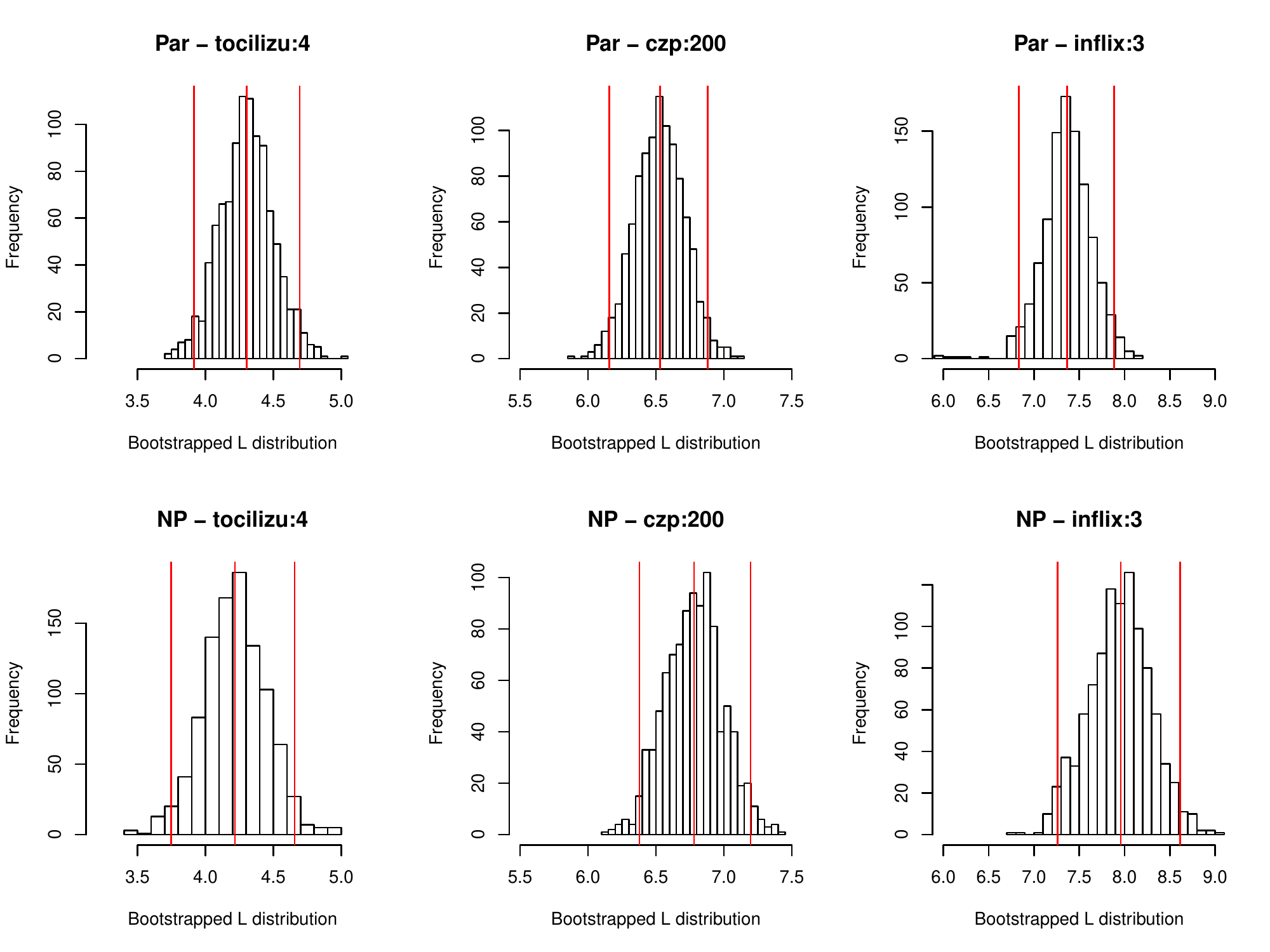}
    \caption{Bootstrapped distribution of L estimates, red lines corresponding to 2.5\%,50\%,97.5\% percentile of the bootstrapped sample. (a = 5, b = 25)} \label{f:bootDist1}
      \includegraphics[width=0.8\linewidth]{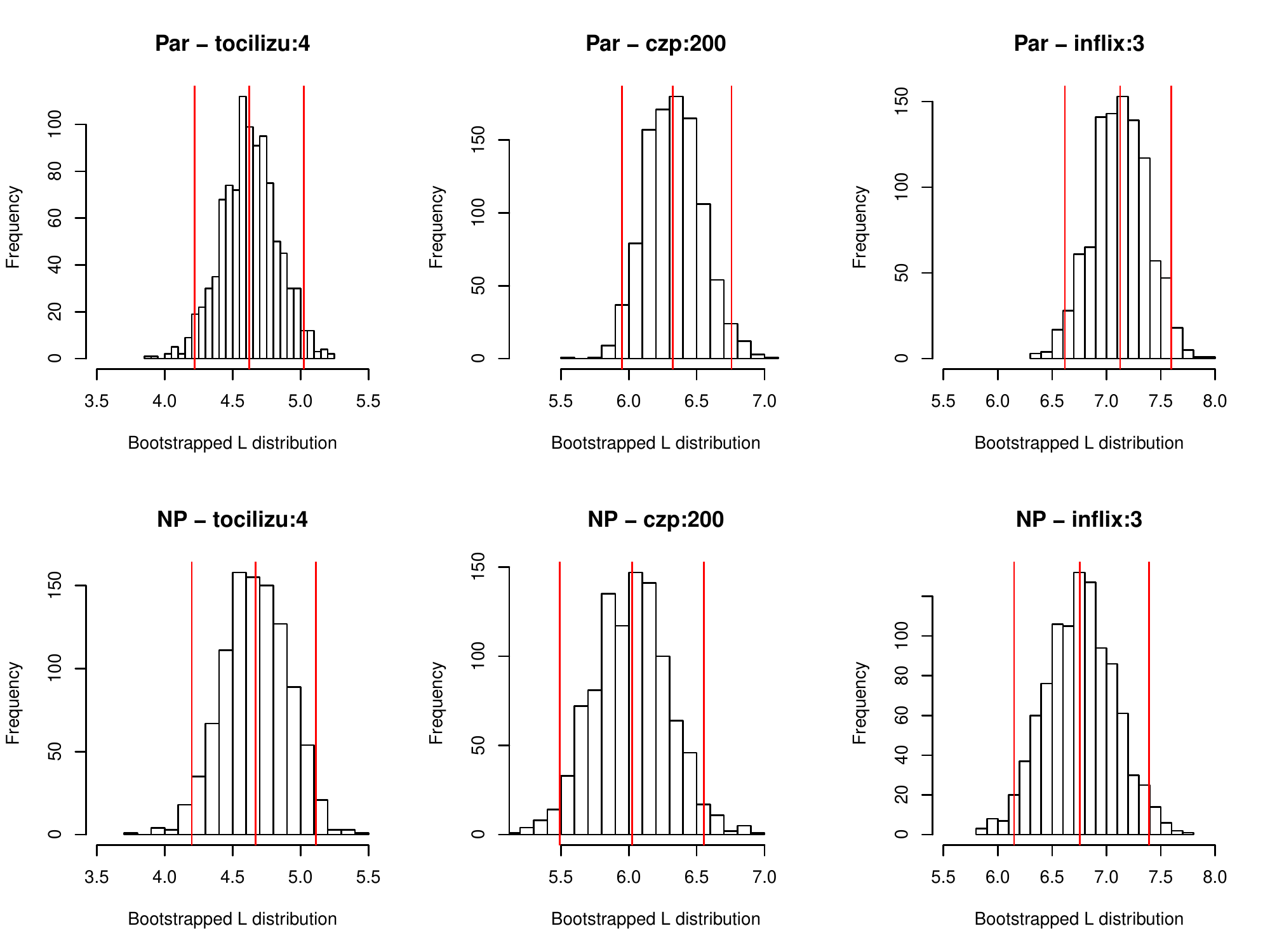}
    \caption{Bootstrapped distribution of L estimates, red lines corresponding to 2.5\%,50\%,97.5\% percentile of the bootstrapped sample. (a = 30, b = 50)} \label{f:bootDist2}
\end{figure}

The selected $M$ values are summarized in Table~\ref{t:realMopt}. Figure~\ref{f:realdata4} shows both parametric and nonparametric fit of the data by treatment, the MTX curves serve as the control treatment for comparison. The grey line corresponding to time interval $a = 5$ and $b = 25$. For demonstration, we also report the results for a longer time interval $a = 30$ and $b = 50$. The estimated $L$-metric quantity and their bootstrapped confidence interval are presented in Table~\ref{t:real_L}. Additionally the entire bootstrapped sampling distribution of the $L$ metric is shown in Figure~\ref{f:bootDist1} and Figure~\ref{f:bootDist2}. 

The bootstrap distributions and associated 95\% confidence intervals reported in the figures and the tables reveal several interesting features and advantages of using a functional metric compared to a fixed time comparison of $\Delta(t)$ at a fixed $t=t_0$. Notice that the 95\% confidence interval of tocilizu:4 has no over with that of czp:200 and inflix:3 (see Table~\ref{t:real_L}) indicating that over the entire time interval $(5, 25)$ weeks, the later two drugs perform significantly better than the first one relative to the baseline treatment MTX. These results are consistent no-matter parametric or nonparametric models are used. On the other hand, the performances of czp:200 and inflix:3 are statistically not different (also see Figure~\ref{f:realdata4}). 

The results for time interval of $(30, 50)$ weeks are similar but they differ by the parametric and nonparametric models. Under the nonparametric model, here appears to be a slight overlap of the 95\% intervals for tocilizu:4 and czp:200, as can be seen in Table~\ref{t:real_L}, but not under parametric model. On the other hand there's no overlap of the 95\% confidence intervals of the $L$ metric when comparing tocilizu:4 and inflix:3, indicating that the later drug perform better even in the long run when the response curves seem to plateau. These findings are slightly different from the parametric model based bootstrapped distributions and associated 95\% confidence intervals. As in our simulation studies we have seen superior and robust performance  of the nonparametric models, we feel more comfortable reporting results best on the nonparametric models although for this case study, the estimated response rate curves are very similar.


\begin{table}[h]
\centering
\begin{tabular}{|r|rr|rr|}
  \hline
 & \multicolumn{2}{c|}{$\hat{\alpha}$} &\multicolumn{2}{c|}{$\hat{\beta}$} \\ 
  \hline
     MTX & 0.156 & (0.008) & 0.398 & (0.084) \\ 
  tocilizu:4 & 0.489 & (0.008) & 0.232 & (0.019) \\ 
  czp:200& 0.575 & (0.008) & 0.435 & (0.033) \\ 
  inflix:3 & 0.614 & (0.011) & 0.549 & (0.126) \\ 
  \hline
\end{tabular}
\caption{RA data: parameter estimate in the exponential model}\label{t:real_par}
\end{table}


\begin{table}[h!]
\centering
\begin{tabular}{|c|c|c|c|}
  \hline
 & tocilizu:4 & czp:200 &  inflix:3 \\
  \hline
$\alpha = 0.2$ & 8  & 9  & 9\\
$\alpha = 0.3$ & 8  & 9  & 9\\
$\alpha = 0.4$ & 9  & 9  & 9\\
  \hline
\end{tabular}
\caption{Selected M values under different critical values after relaxing the monotonic constraint.}\label{t:realMopt}
\end{table}

\begin{table}[ht]
\centering
\small
\begin{tabular}{|c|cc|cc|}
		\hline
				& \multicolumn{2}{c|}{Parametric} & \multicolumn{2}{c|}{Nonparametric}\\
		\hline
	& $\hat{L}$ & $95\%$ CI &$\hat{L}$  & $95\%$ CI\\ 
		\hline
	 &\multicolumn{4}{c|}{a = 5,  b = 25}\\ 
		\hline
tocilizu:4 & 4.303 (0.187)  & (3.978,4.605) & 4.200 (0.233) & (3.830, 4.597) \\ 
czp:200 & 6.516 (0.183) & (6.230, 6.835) & 6.722(0.220) & (6.423 ,7.135) \\ 
inflix:3 & 7.385 (0.275) & (6.911,7.769) & 7.893 (0.334)  & (7.377,8.477) \\ 
		\hline
	& \multicolumn{4}{c|}{a = 30,  b = 50}\\ 
  \hline
tocilizu:4 & 4.620 (0.207) & (4.273, 4.974) & 4.649 (0.238 )& (4.268, 5.054)\\ 
czp:200  & 6.330 (0.206) & (6.002, 6.668) & 6.003 (0.279) &  (5.561, 6.463)\\ 
inflix:3 & 7.120 (0.255) & (6.693, 7.548) & 6.699 (0.318) &  (6.248, 7.286)\\ 
  \hline
\end{tabular}
\caption{Functional Metric $L$ estimates (bootstrapped s.e.) and 95\% confidence interval.}
\label{t:real_L}
\end{table}

\section{Conclusions and Discussions}
\label{conc}

In this paper, we present a functional metric to assess biosimilarity. Functional metrics provide stronger evidence in support of or against when comparing two drugs. We also compare parametric models verse nonparametric model. We show in simulation studies that nonparametric approach provides more flexible model forms and are robust against model misspecification. In our case study with RA trials, we have shown the added utility of using functional metric over fixed time point difference of response rates. 

Recall that in our models, we have considered $Y_j(t) \sim \mathrm{Bin}\{n_j,\theta_j(t)\}, \text{for }t\ge0$ (as in (\ref{eq:y})). As an extension of the models, it is possible to allow the number of observations $n_j$ to change over time $t$ as well. To enable that, we denote $S_j(t)$ to be the number of patients that respond at time $t$ in the $j$th treatment group. We can then use the following hierarchical model:
$$
S_j(t) \sim \mathrm{Poisson}\left(\lambda_j(t)\right),
$$
and 
$$
Y_j(t)\vert S_j(t) \sim \mathrm{Bin}\left(S_j(t),\theta_j(t)\right).
$$
It would be interesting to develop estimation methods for the above extended hierarchical models using both parametric and nonparametric models. 

\bibliographystyle{chicago}
\bibliography{citations.bib}

\newpage
\begin{appendices}
\section*{Appendix: Random coefficients model for MTX}
\begin{figure}[h]
\centering
  \includegraphics[width=0.6\linewidth]{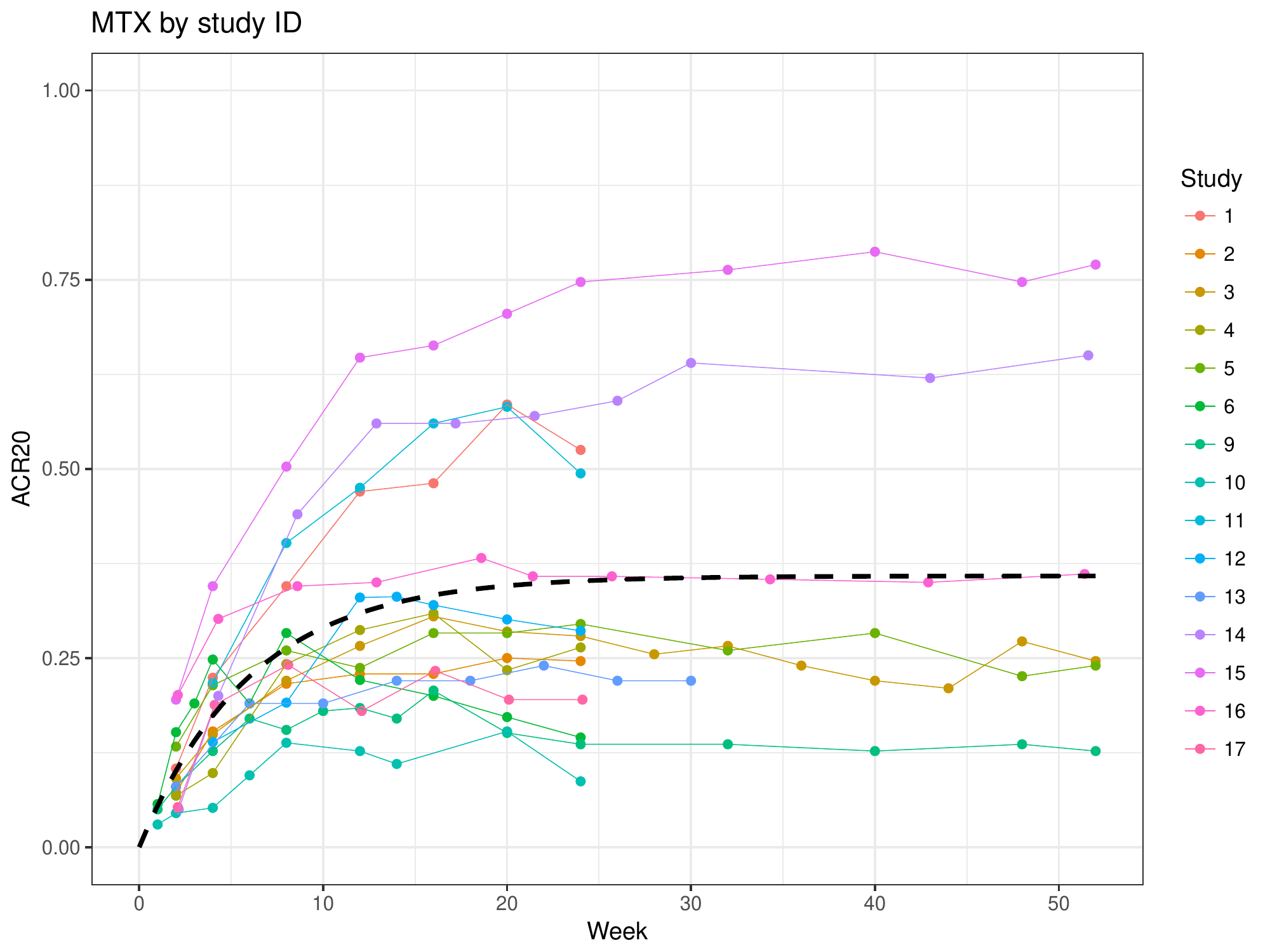}
    \caption{MTX data by study ID. The dotted line is the fitted line using all the data under fixed effect model.}\label{f:MTX}
\end{figure}

We notice that the baseline treatment MTX demonstrates a notable variation across different studies in the RA data set, as shown in Figure~\ref{f:MTX}. Therefore, a random coefficients model could be a good choice to fit the MTX data.
\begin{align*}
    Y_{ij} \mid \alpha_i, \beta_i &\sim \mathrm{Bin}[n_{ij},\theta(t_{ij};\alpha_i,\beta_i)]\\
    \begin{pmatrix}
    \alpha_i\\
    \beta_i
    \end{pmatrix}\mid \mu_a,\mu_b,\sigma_a,\sigma_b,\sigma_{ab} &\sim \mathrm{lognormal}\left[
    \begin{pmatrix}
    \mu_a\\
    \mu_b
    \end{pmatrix},\begin{pmatrix}
    \sigma^2_a, \sigma_{ab}\\
    \sigma_{ab}, \sigma^2_b
    \end{pmatrix}\right],
\end{align*}
where $\theta(t_{ij};\alpha_i,\beta_i) = \alpha_i(1 - e^{-\beta_i  t_{ij}})$.
Denote $\eta = (\mu_a,\mu_b,\sigma_a,\sigma_b,\sigma_{ab})^T$.
We could obtain the maximum likelihood estimator of $\eta$ by a two step procedure.

\begin{figure}[h]
\centering
  \includegraphics[width=0.6\linewidth]{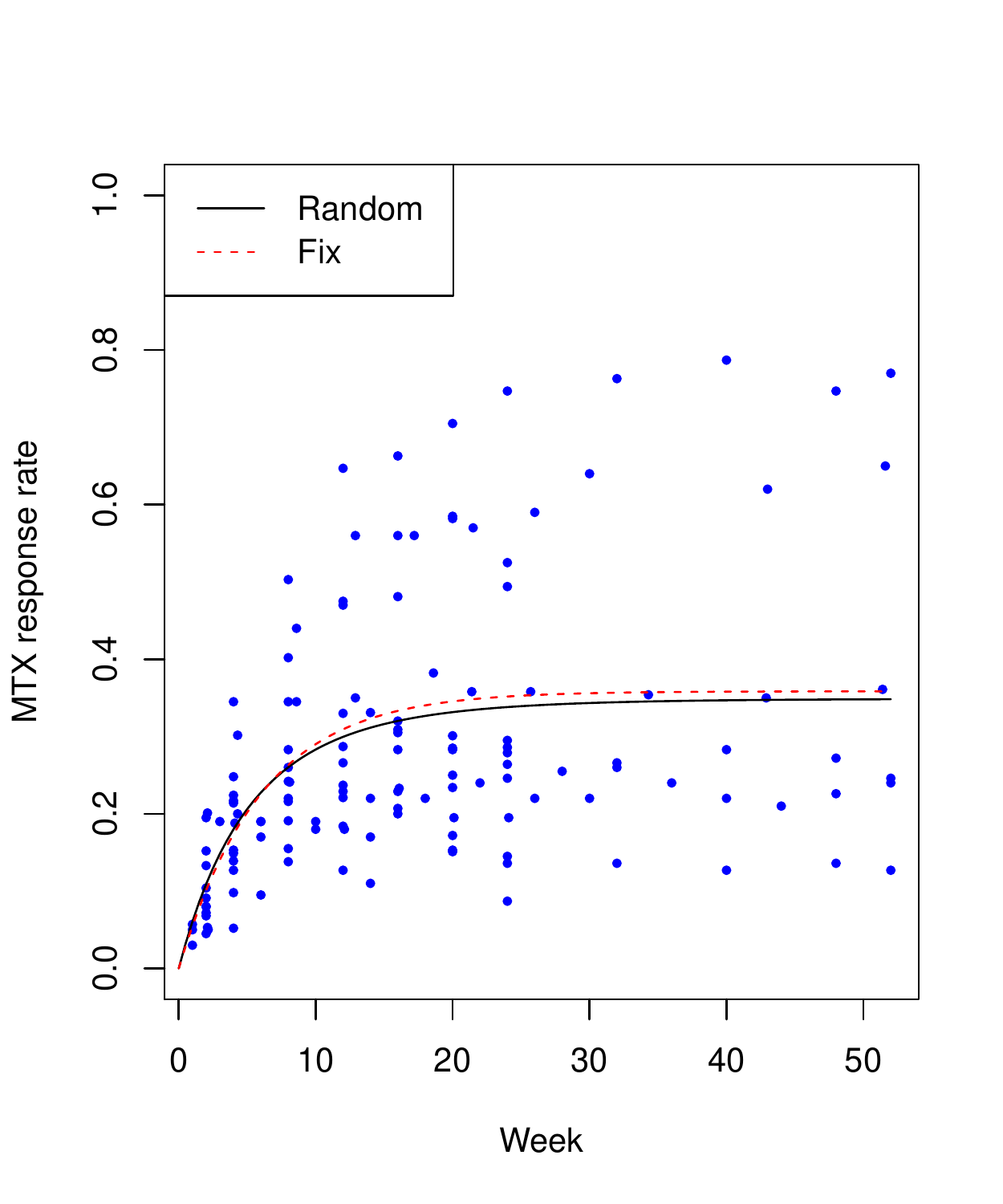}
    \caption{MTX: Comparison of fixed and random coefficient model.}\label{f:MTX_fVSr}
\end{figure}

The the expected value of $\theta(t_{ij})$ is
\begin{align}
    E[\theta(t_{ij};\alpha_i,\beta_i)] &= E[ \alpha_i (1 - e^{-\beta_i t_{ij}}]\nonumber\\
    & = E(\alpha_i) - E(\alpha_i e^{-\beta_i t_{ij}})\label{e:condExp}
\end{align}
We have 
\begin{align*}
    \mathrm{log}\alpha_i \sim N(\mu_a,\sigma^2_a), &\quad \mathrm{log}\beta_i \sim N(\mu_b,\sigma^2_b)\\
    \mathrm{log}\alpha_i \mid \mathrm{log}\beta_i &\sim N(\mu_{a\mid b}, \sigma^2_{a\mid b}),
\end{align*}
where $\mu_{a \mid b} = \mu_a + \frac{\sigma_{ab}}{\sigma^2_b}(\mathrm{log}\beta_i - \mu_b)$ and $\sigma_{a\mid b}^2 = \sigma^2_a - (\sigma_{ab}/\sigma_b)^2$.
Therefore the first term on the RHS of  (\ref{e:condExp}) is $E(\alpha_i) = E(e^{\mathrm{log}\alpha_i}) = e^{\mu_a + \sigma_a^2/2}$. The second term on the RHS of (\ref{e:condExp}) can be write as
\begin{align}
    E(\alpha_i e^{-\beta_i t_{ij}}) & = E\left[ E(\alpha_i e^{-\beta_i t_{ij}}\mid \beta_i) \right]\nonumber\\
    & = E\left[e^{-\beta_i t_{ij}}E(e^{\mathrm{log}\alpha_i}\mid \mathrm{log}\beta_i) \right]\nonumber\\
    & = \mathrm{exp}\left\{\mu_a - \frac{\sigma_{ab}}{\sigma_b^2}\mu_b + \frac{1}{2}\left[\sigma^2_a - \left(\frac{\sigma_{ab}}{\sigma_b}\right)^2\right]\right\}E\left[\mathrm{exp}\left(-\beta_i t_{ij} + \frac{\sigma_{ab}}{\sigma_b^2}\mathrm{log}\beta_i\right)\right].\label{e:CPLXexpectation}
\end{align}
The expectation in the RHS of (\ref{e:CPLXexpectation}) can be approximated via numerical integration. 
The estimates of the random coefficients are summarized in Table~\ref{t:randomCoef}. A comparison of the random coefficients and fixed coefficients models is shown in Figure~\ref{f:MTX_fVSr}.
\begin{table}[h]
\centering
\begin{tabular}{rrrrrr}
  \hline
 &$\hat{\mu}_a$ & $\hat{\mu}_b$ & $\hat\sigma_a$& $\hat{\sigma}_b$ & $\hat{\sigma}_{ab}$ \\
 \hline
Estimate & -1.191 & -1.55  & 0.525  & 0.425 & -0.180 \\ 
s.e. &  0.137 &  0.124 &  0.098  & 0.096 &  0.080\\
\hline
\end{tabular}
\caption{MTX: parameter estimates (MLE) of the random coefficients model}\label{t:randomCoef}
\end{table}

\end{appendices}

\end{document}